\documentclass[
    ,final            
  ]
  {aipproc}

\layoutstyle{6x9}

\usepackage{amssymb}
\usepackage{amsmath}
\usepackage{color}
\begin{document}

\newcommand{\pen}{$\pi$$\to$$\text{e}$$\nu$}
\newcommand{\peII}{$\pi_{\text{e}2}$}
\newcommand{\pme}{$\pi$$\to$$\mu$$\to$$\text{e}$}
\renewcommand{\today}{27 August 2012}

\title{New studies of allowed pion and muon decays}

\newcommand{\uva}{Institute of Nuclear and Particle Physics, University
             of Virginia,  Charlottesville, VA 22904, USA} 
\newcommand{\dubna}{Joint Institute for Nuclear Research, RU-141980 Dubna,
                 Russia} 
\newcommand{\PSI}{Paul Scherrer Institut, Villigen PSI, CH-5232, Switzerland}
\newcommand{\IRB}{Institut Rudjer Bo\v{s}kovi\'c, HR-10000 Zagreb,
                 Croatia} 
\newcommand{\swierk}{NCBJ National Centre for Nuclear Research,
                  Otwock, Poland} 
\newcommand{\tbilisi}{Institute for High Energy Physics, Tbilisi State
                 University, GUS-380086 Tbilisi, Georgia} 
\newcommand{\unizh}{Physik-Institut, Universit\"at Z\"urich, CH-8057
                 Z\"urich, Switzerland}

\author{D.~Po\v{c}ani\'c}{address=\uva}
\author{A.~Palladino}{address=\uva,altaddress=\PSI}
\author{L.~P.~Alonzi}{address=\uva}
\author{V.~A.~Baranov}{address=\dubna}
\author{W.~Bertl}{address=\PSI}
\author{M.~Bychkov}{address=\uva}
\author{Yu.M.~Bystritsky}{address=\dubna}
\author{E.~Frle\v{z}}{address=\uva}
\author{V.A.~Kalinnikov}{address=\dubna}
\author{N.V.~Khomutov}{address=\dubna}
\author{A.S.~Korenchenko}{address=\dubna}
\author{S.M.~Korenchenko}{address=\dubna}
\author{M.~Korolija}{address=\IRB}
\author{T.~Kozlowski}{address=\swierk}
\author{N.P.~Kravchuk}{address=\dubna}
\author{N.A.~Kuchinsky}{address=\dubna}
\author{M.C.~Lehman}{address=\uva}
\author{D.~Mekterovi\'c}{address=\IRB}
\author{E.~Munyangabe}{address=\uva}
\author{D.~Mzhavia}{address=\dubna,altaddress=\tbilisi}
\author{P.~Robmann}{address=\unizh}
\author{A.M.~Rozhdestvensky}{address=\dubna}
\author{S.N.~Shkarovskiy}{address=\dubna}
\author{U.~Straumann}{address=\unizh}
\author{I.~Supek}{address=\IRB}
\author{P.~Tru\"ol}{address=\unizh}
\author{Z.~Tsamalaidze}{address=\tbilisi}
\author{A.~van~der~Schaaf}{address=\unizh}
\author{E.P.~Velicheva}{address=\dubna}
\author{V.P.~Volnykh}{address=\dubna}

\begin{abstract}
Building on the rare pion and muon decay results of the PIBETA
experiment, the PEN collaboration has undertaken a precise measurement
of $B_{\pi\text{e}2}\equiv R^\pi_{\text{e}/\mu}$, the $\pi^+ \to
\text{e}^+\nu(\gamma)$ decay branching ratio, at the Paul Scherrer
Institute, to reduce the present 40$\times$ experimental precision lag
behind theory to $\sim 6-7\times$.
Because of large helicity suppression, $R^\pi_{\text{e}/\mu}$ is
uniquely sensitive to contributions from non-$(V-A)$ physics, making
this decay a particularly suitable subject of study.  Even at current
precision, the experimental value of $B_{\pi\text{e}2}$ provides
the most accurate test of lepton universality available.
During runs in 2008--10, PEN has accumulated over $2\times 10^7$
\peII\ events; a comprehensive maximum-likelihood analysis is 
currently under way.  The new data will also lead to improved
precision of the earlier PIBETA results on radiative $\pi$ and $\mu$
decays.
\end{abstract}

\maketitle



Historically, the \pen\ (or \peII) decay, provided an early
confirmation of the $V-A$ nature of the electroweak interaction.
Thanks to exceptionally well controlled theoretical uncertainties, its
branching ratio is now understood at better than a part in $10^4$.
The most recent independent theoretical calculations are in excellent
agreement, and give:
\begin{equation}
     B_{\pi\text{e}2}^{\text{SM}} \equiv R_{\text{e}/\mu}^{\pi,\text{SM}} =
       \frac{\Gamma(\pi \to \text{e}\bar{\nu}(\gamma))}
          {\Gamma(\pi \to  \mu\bar{\nu}(\gamma))}\bigg|_{\text{calc}} =
   \begin{cases}
     1.2352(5) \times 10^{-4} & \text{Ref.~\cite{Mar93},} \\
     1.2354(2) \times 10^{-4} & \text{Ref.~\cite{Fin96},} \\
     1.2352(1) \times 10^{-4} & \text{Ref.~\cite{Cir07},} \\
   \end{cases}
\end{equation}
where ``$(\gamma)$'' indicates that radiative decays are included.
Marciano and Sirlin \cite{Mar93} and Finkemeier \cite{Fin96} took into
account radiative corrections, higher order electroweak leading
logarithms, short-distance QCD corrections, and structure-dependent
effects, while Cirigliano and Rosell \cite{Cir07} used two-loop chiral
perturbation theory.  A number of exotic processes outside of the
current standard model (SM) can produce deviations from the above
predictions based on lepton universality, mainly through induced
pseudoscalar (PS) currents.  Prime examples are: charged Higgs in
theories with multiple Higgs bosons, PS leptoquarks in theories with
dynamical symmetry breaking, classes of vector leptoquarks, parameters
of certain SUSY partner particles, as well as non-zero neutrino masses
and their mixing (Refs.~\cite{Rai08,Bry11} give recent reviews of the
subject).  Thus, \peII\ decay complements direct searches for new
physics at modern colliders.

The two most recent measurements of the branching ratio
\cite{Bri92,Cza93} are mutually consistent and dominate the world
average of $1.230(4) \times 10^{-4}$, which, however, trails the
theoretical accuracy by a factor of 40.  The PEN experiment
\cite{pen06} is aiming to reach ($\Delta B/B)_{\pi\text{e}2} \simeq
5\times 10^{-4}$, and in doing so, to set new limits on the above
non-SM processes.  PEN also aims to improve the PIBETA results for
radiative decays $\pi^+\to \text{e}^+\nu\gamma$ and $\mu^+\to
\text{e}^+\nu\bar{\nu}\gamma$.  Meanwhile, PiENu \cite{PiENu}, a
complementary experiment currently under way at TRIUMF, has a similar
goal for $(\Delta B/B)_{\pi\text{e}2}$.

\begin{figure}[tb]
  \parbox{0.56\linewidth}{
     \includegraphics[width=\linewidth]{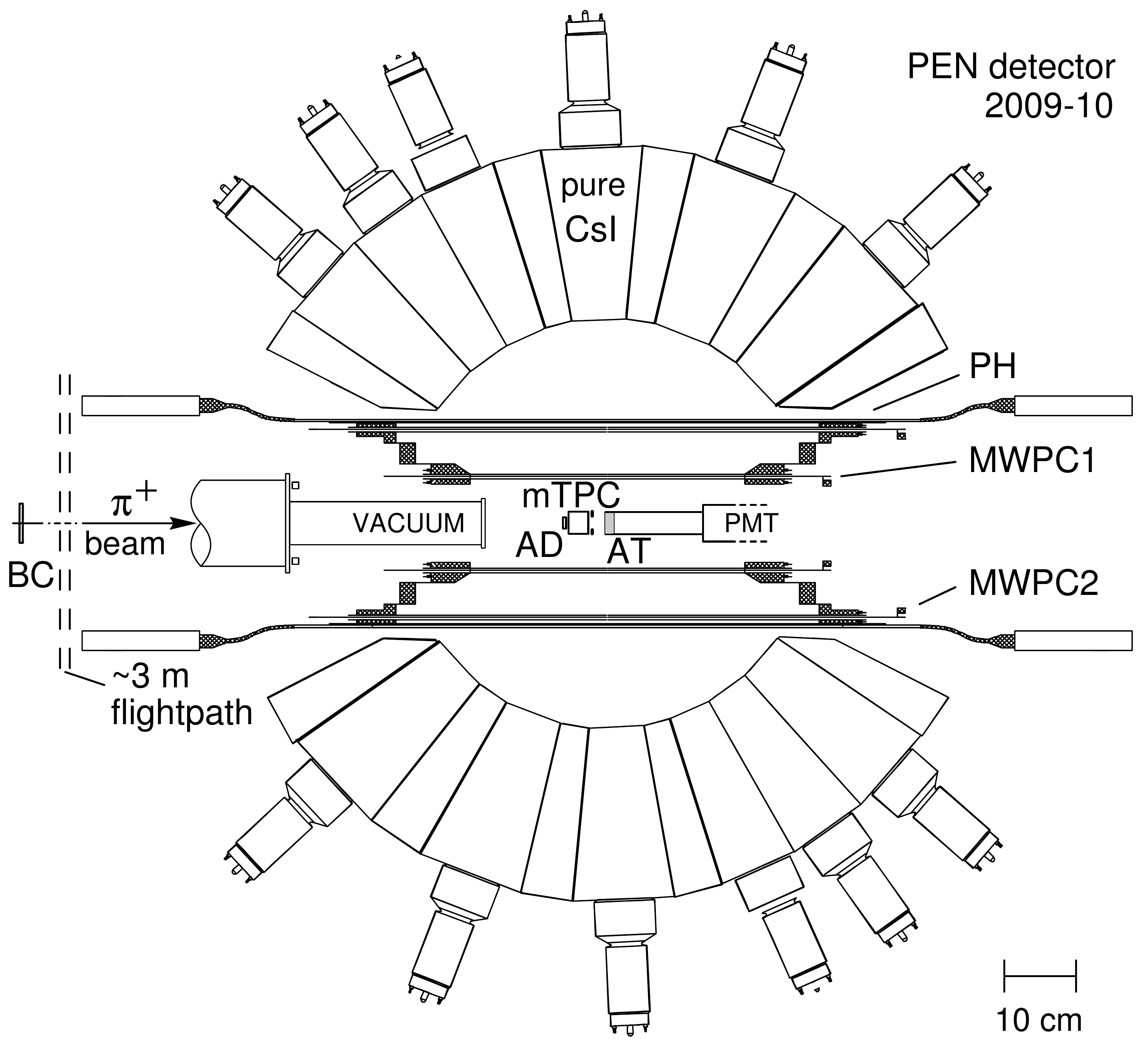} \\[-12pt]
     \begin{picture}(0,0)
       \put (10,200) {(a)}
     \end{picture}
                         }
  \hspace*{\fill}
  \parbox{0.43\linewidth} {\hspace*{\fill}
     \includegraphics[width=0.85\linewidth,height=0.15\textheight]
                                  {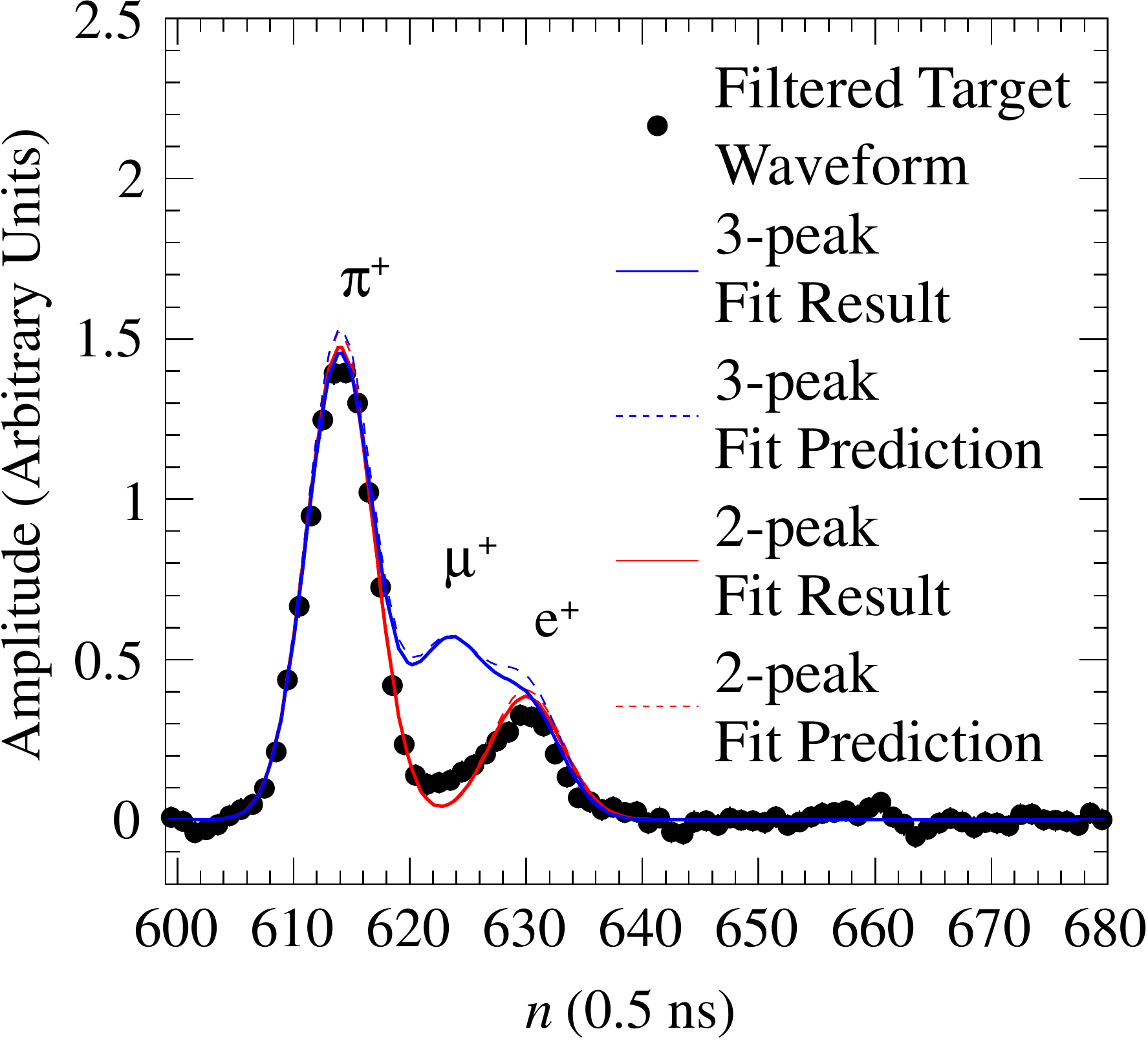}  \hspace*{\fill}\\
     \includegraphics[width=\linewidth]{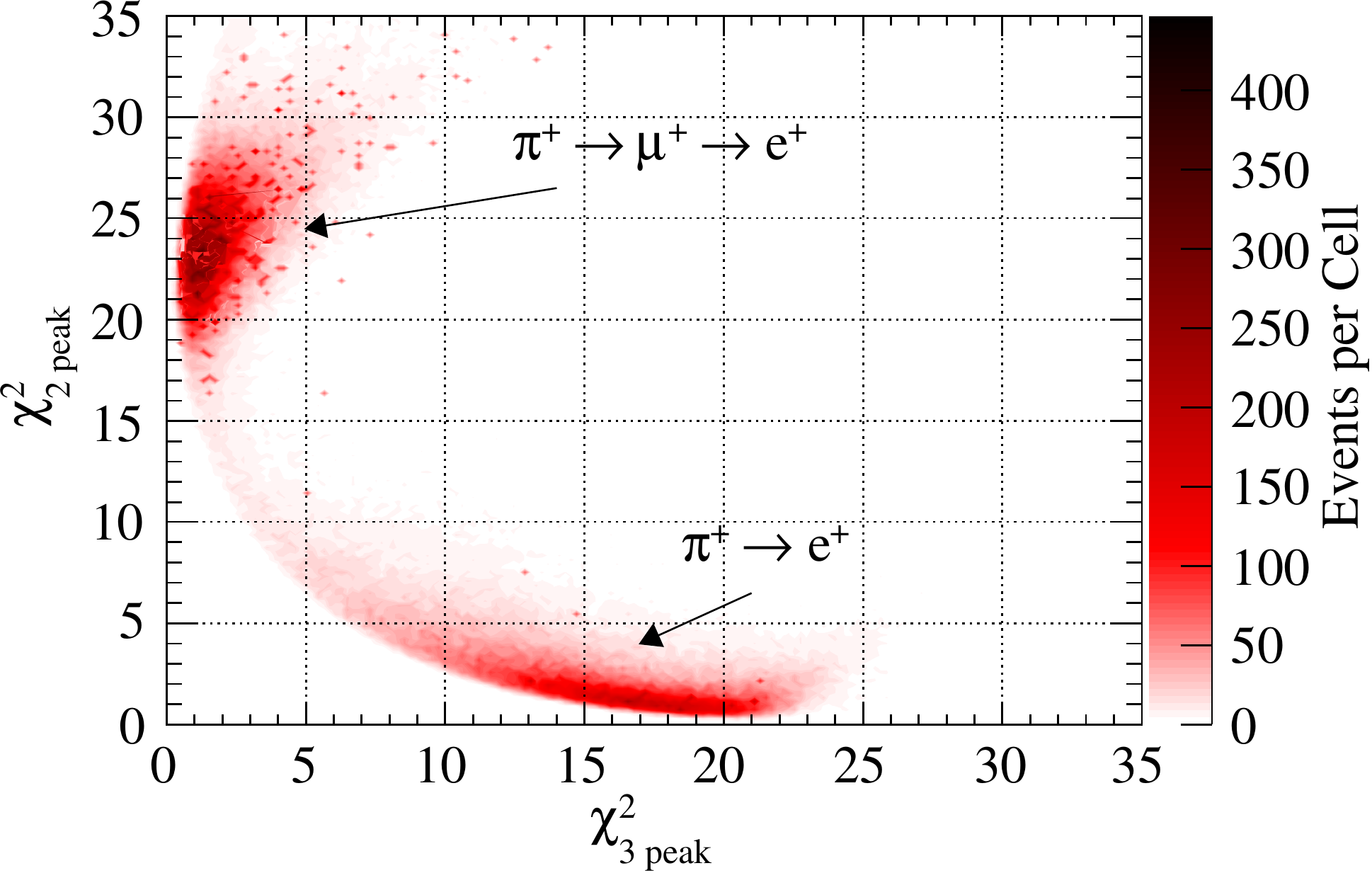} 
     \vspace*{-\baselineskip}
      \begin{picture}(0,0)
        \put (45,205) {(b)}
        \put (130,105) {(c)}
      \end{picture}
                          }
    \caption{(a) Cross section drawing of the PEN detector system
      (details in text).
      (b) Filtered active target  waveform for a \pen\ event
      showing a strongly preferred 2-peak fit.
      (c) Excellent $\chi^2$ separation of 2- and 3-peak filtered
      waveform fits (bottom). 
      \label{fig:pen_det} }
\end{figure}
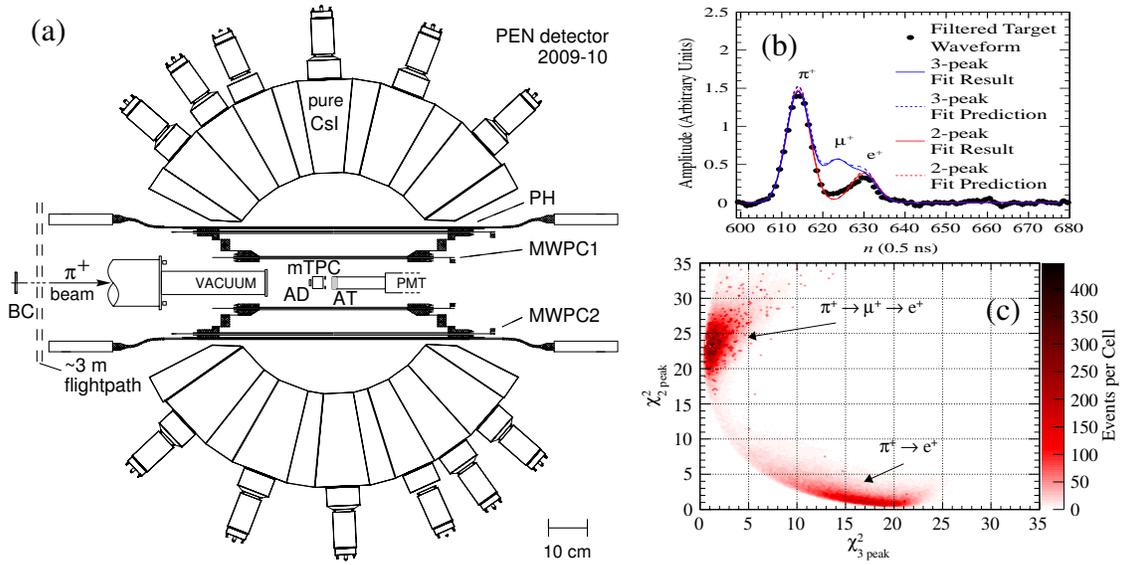

The PEN experiment uses an upgraded version of the PIBETA detector
system, described in detail in Ref.~\cite{Frl04a}, and used in a
series of rare pion and muon decay measurements \cite{Poc04, Frl04b,
  Byc09}.  The PEN apparatus, shown in Fig.~\ref{fig:pen_det}(a),
consists of a large-acceptance ($\sim 3\pi$\,sr) electromagnetic
shower calorimeter (pure CsI, 12 radiation lengths thick) with
non-magnetic tracking in concentric cylindrical multi-wire
proportional chambers (MWPC1,2) and plastic scintillator hodoscope
(PH), surrounding a plastic scintillator active target (AT).  Beam
pions pass through an upstream detector (BC), lose energy in the
active degrader (AD), are tracked in a mini time projection chamber
(mTPC), and stop in the AT.  Signals from the beam detectors are
digitized in waveform digitizers, running at 2\,GS/s for BC, AD, and
AT, and at 250\,MS/s for mTPC.
\\
\indent
The \peII\ branching ratio will be evaluated by normalizing the
observed yield of \pen\ decays to the number of sequential decays
\pme, within a 250\,ns gate starting some 40\,ns before the pion stop
time \cite{pen06}.  Assignment of detected events to either of the two
processes, or to a background process, is made within a comprehensive,
blind and unbinned maximum likelihood analysis (MLA) \cite{Pal12}.
Key to achieving the goal uncertainty is in the control of the
systematics.

One among many powerful tools available to develop reliable
probability density functions for the MLA is provided by the beam
counter waveform (wf) digitizer data, helping us to obtain clean
samples of \peII\ events (two pulses in the AT-wf) and sequential
\pme\ decay events (three pulses in the AT-wf).  The procedure and
results are summarily illustrated in Fig.~\ref{fig:pen_det}(b,c).
Another key component in the analysis is the comprehensive GEANT4
Monte Carlo simulation of the experiment that produces synthetic data
fully equivalent to the measured data \cite{Alo12}.

Furthermore, thanks to lower and better controlled backgrounds, the
PEN data will enable us to improve on the precision of the previous
PIBETA results on pion \cite{Byc09} and muon \cite{Van06,Mun12}
radiative decays, both of which are sensitive to non-$(V-A)$
interactions.  Besides providing the most stringent limit on tensor
interactions to date \cite{Bha12}, the PIBETA
$\pi\to\text{e}\nu\gamma$ results \cite{Byc09} also furnish
fundamental inputs for chiral perturbation theory related to pion
structure.  On the other hand, muon radiative decay, being largely
free from hadronic corrections, provides model-independent information
on non-$(V-A)$ admixtures in the lagrangian.

Three PEN data runs have been completed, in 2008, 2009, and 2010,
respectively, collecting over 20\,M \peII\ events.  Comprehensive data
analysis, focusing on control of systematics, is currently under way.

%
%

\medskip
  This work has been supported by grants from the US National Science
  Foundation (most recently PHY-0970013), the Paul Scherrer Institute,
  and the Russian Foundation for Basic Research (Grant 08-02-00652a).

%

\begin{thebibliography}{00}
%
%
%
%
\bibitem{Mar93} W.J. Marciano and A. Sirlin,
    Phys.\ Rev.\ Lett.\ {\bfseries 71} (1993) 3629.

\bibitem{Fin96} M. Finkemeier, Phys.\ Lett.\ B {\bfseries 387} (1996)
  391. 

\bibitem{Cir07} V. Cirigliano and I. Rosell, Phys.\ Rev.\ Lett.\
  {\bfseries 99} (2007) 231801.

\bibitem{Rai08} M. Raidal, A. van der Schaaf, I. Bigi, et al., Eur.\
  Phys.\ J C {\bfseries 57} (2008) 13.

\bibitem{Bry11} D. Bryman, W.J. Marciano, R. Tschirhart, T. Yamanaka,
  Ann.\ Rev.\ Nucl.\ Part.\ Sci.\ \textbf{61} (2011) 331. 

\bibitem{Bri92} D.I. Britton, S. Ahmad, D.A. Bryman, et al., Phys.\
  Rev.\ Lett.\ {\bfseries 68} (1992) 3000.

\bibitem{Cza93} G. Czapek, A. Federspiel, A. Fl\"ukiger, et. al.,
  Phys.\ Rev.\ Lett.\ {\bfseries 70} (1993) 17. 

\bibitem{pen06} ``A precise measurement of the $\pi^+$$\to$$e^+\nu$
  branching ratio,'' PSI experiment proposal R-05-01, 
  \mbox{\url{http://pen.phys.virginia.edu/proposal/pen_proposal.pdf}}
  (2006).  

\bibitem{PiENu} PiENu: \url{http://pienu.triumf.ca}, and A. Sher,
  elsewhere in these Proceedings. 

\bibitem{Frl04a} E. Frle\v{z}, D. Po\v{c}ani\'c, K. Assamagan, et al.,
  Nucl.\ Inst.\ and Meth.\ A {\bfseries 526} (2004) 300.

\bibitem{Poc04} D. Po\v{c}ani\'c, E. Frle\v{z}, V.A. Baranov, et al.,
  Phys.\ Rev.\ Lett.\ {\bfseries 93} (2004) 181803.

\bibitem{Frl04b} E. Frle\v{z}, D. Po\v{c}ani\'c, V.A. Baranov, et al.,
  Phys.\ Rev.\ Lett.\ {\bfseries 93} (2004) 181804.

\bibitem{Byc09} M. Bychkov, D. Po\v{c}ani\'c, B.A. VanDevender, et
  al., Phys.\ Rev.\ Lett.\ \textbf{103} (2009) 051802.

\bibitem{Pal12} A. Palladino, PhD thesis, University of Virginia
  (2012). 

\bibitem{Alo12} L.P. Alonzi, PhD thesis, University of Virginia
  (2012).

\bibitem{Van06} B.A. VanDevender, PhD thesis, University of Virginia
  (2006). 

\bibitem{Mun12} E. Munyangabe, PhD thesis, University of Virginia
  (2012), and article in preparation. 

\bibitem{Bha12} T. Bhattacharya, V. Cirigliano, S.D. Cohen, et al.,
  Phys.\ Rev.\ D \textbf{85} (2012) 054512.

%
\end{thebibliography}
\end{document}